\documentclass[letterpaper]{article} 
\usepackage{aaai2026}  
\usepackage{times}  
\usepackage{helvet}  
\usepackage{courier}  
\usepackage[hyphens]{url}  
\usepackage{graphicx} 
\usepackage{natbib}  
\setcitestyle{numbers}
\usepackage{caption} 
\usepackage{algorithm}
\usepackage{algorithmic}
\usepackage{amsmath}
\usepackage{amssymb}

\usepackage{xcolor}  

\usepackage{newfloat}
\usepackage{listings}
\DeclareCaptionStyle{ruled}{labelfont=normalfont,labelsep=colon,strut=off} 
\floatstyle{ruled}
\newfloat{listing}{tb}{lst}{}
\floatname{listing}{Listing}
\title{AgentTrace: A Structured Logging Framework for Agent System Observability}

\author {
    Adam AlSayyad\thanks{Equal contribution.},
    Kelvin Yuxiang Huang\footnotemark[1],
    Richik Pal\footnotemark[1]
}

\affiliations {
    University of California, Berkeley\\
    alsayyad@berkeley.edu, yxkelvinhuang@berkeley.edu, richik.pal@berkeley.edu
}

\begin{document}

\maketitle

\begin{abstract}
Despite the growing capabilities of autonomous agents powered by large language models (LLMs), their adoption in high-stakes domains remains limited. A key barrier is security: the inherently nondeterministic behavior of LLM agents defies static auditing approaches that have historically underpinned software assurance. Existing security methods, such as proxy-level input filtering and model glassboxing, fail to provide sufficient transparency or traceability into agent reasoning, state changes, or environmental interactions. In this work, we introduce AgentTrace, a dynamic observability and telemetry framework designed to fill this gap. AgentTrace instruments agents at runtime with minimal overhead, capturing a rich stream of structured logs across three surfaces: operational, cognitive, and contextual. Unlike traditional logging systems, AgentTrace emphasizes continuous, introspectable trace capture, designed not just for debugging or benchmarking, but as a foundational layer for agent security, accountability, and real-time monitoring. Our research highlights how AgentTrace can enable more reliable agent deployment, fine-grained risk analysis, and informed trust calibration, thereby addressing critical concerns that have so far limited the use of LLM agents in sensitive environments.
\end{abstract}

\section{Introduction}

The deployment of autonomous agents built on large language models (LLMs) has shown early promise across a wide spectrum of domains, including software engineering, scientific analysis, and complex decision-making workflows \cite{schick2023toolformer, weng2023autonomous}. However, despite the functional competence of these systems in isolated tasks, their adoption in safety-critical or high-integrity environments remains severely limited. A primary constraint is the absence of structured and dynamic observability frameworks that can account for the stochastic reasoning behaviors of LLM agents and enable reliable diagnosis, security assessment, and governance.

Conventional methods for securing AI systems, such as static input filtering, prompt hardening, and API boundary control, are insufficient for LLM-based agents that act through long-running, multi-step reasoning cycles in open-ended environments. These agents dynamically compose tool invocations, retrieve external knowledge, and revise goals during execution, producing behaviors that are difficult to trace or explain post hoc. Current security and auditing tools remain constrained by assumptions of determinism, procedural transparency, or bounded action space, assumptions that do not hold in LLM-based settings. As a result, the dominant paradigm has centered on proxy-level defenses (e.g., PromptArmor) and glassbox introspection of static prompts and outputs \cite{zou2023universal}. These approaches offer limited insight into agent intent, decision provenance, or operational context, especially in the presence of tool-use chaining, memory operations, or multi-agent collaboration.

Crucially, the undeterministic nature of agentic reasoning introduces a novel challenge for security and governance: threats and failures can emerge not merely from malicious inputs or faulty tools, but from the emergent behavior of the agent's cognitive trajectory. As observed in recent red-teaming and adversarial prompting research \cite{liu2024agentpatterns, zou2023universal}, even well-scoped agents may deviate from expectations when their reasoning states are not explicitly monitored. To address this, a shift is required from static, perimeter-oriented security architectures toward dynamic, semantic observability of the agent's internal execution process.

Our contributions are as follows. We present \textbf{AgentTrace}, a structured, schema-based logging framework that instruments LLM agents at runtime without requiring code modifications. We introduce a three-surface taxonomy: cognitive, operational, and contextual, that enables multi-level introspection into an agent’s reasoning, execution, and environment. Finally, we demonstrate how AgentTrace integrates with existing telemetry infrastructures such as OpenTelemetry to provide scalable, real-time observability. Through this design, AgentTrace establishes a foundation for transparent, accountable, and reproducible agentic systems, paving the way for future research in alignment, evaluation, and safety.

\section{Related Work}
\paragraph{Agent Observability and Tracing Frameworks.} Recent agent-oriented observability tools instrument execution flows to support debugging and monitoring. \textit{AgentOps} introduces a hierarchical span taxonomy that organizes reasoning, planning, workflow, task, tool, and LLM spans to trace artifacts and processes throughout the agent lifecycle \cite{dong2024agentopsenablingobservabilityllm}. \textit{LADYBUG} complements this with post-hoc debugging that combines execution tracing and LLM-aided self-reflection \cite{rorseth2025ladybug}. However, these systems primarily target \emph{single-surface} traces and lack a schema that unifies cognitive artifacts with operational and contextual signals. In contrast, AgentTrace introduces a \emph{schema-based}, \emph{multi-surface} observability model linking \emph{operational}, \emph{cognitive}, and \emph{contextual} traces under a unified envelope, realized at runtime via lightweight instrumentation. 

\paragraph{System-Level Telemetry and Distributed Tracing.} System-centric approaches provide horizontal visibility into API calls and service dependencies, often via kernel/OS boundary tracing and OpenTelemetry-style pipelines. \textit{AgentSight}, for example, correlates LLM prompts with kernel events using eBPF to bridge intent and execution at system boundaries, and has been applied to boundary tracing and anomaly detection \cite{Zheng_2025}. Yet these methods are largely semantics-agnostic to agent intent and internal reasoning, offering limited \emph{causal} linkage between what the agent infers and what the system executes. \textbf{AgentTrace} complements them by \emph{embedding cognitive semantics into the telemetry stream}: cognitive spans are \emph{nested} within operational and contextual spans and exported through standard backends, preserving interoperability while enabling reasoning-aware, end-to-end traces. 

\paragraph{Cognitive Interpretability and Textual Traceability.} Work on cognitive observability and agentic interpretability models reasoning traces and human–agent alignment. \textit{Watson} surfaces implicit reasoning errors in LLM-powered agents without altering agent architecture \cite{rombaut2024watsoncognitiveobservabilityframework}, while concurrent work frames explanation as interactive mental-model building with LLM-driven proactive clarification \cite{kim2025llmspursueagenticinterpretability}. These efforts enhance understanding of internal reasoning but remain decoupled from runtime observability and structured, composable telemetry. \textbf{AgentTrace} bridges this gap by treating cognition as a first-class telemetry surface: reasoning steps, plans, and reflections are captured in a machine-readable schema and \emph{causally linked} to operational actions and contextual I/O at runtime, enabling unified, schema-consistent analysis across surfaces.”

\section{Methodology}
\label{sec:Methodology}

\subsection{Schema}
We present a principled, schema-based methodology for capturing rich, interpretable traces of autonomous LLM-agent behavior. At the core of our logging framework is a formalized representation of logs as transformations of runtime events into structured records:
\[
L(S\!:\!E\!:\!C)\,\to\,R,
\]
where S denotes the surface (cognitive, operational, or contextual), E is the event content, C represents metadata context, and R is a structured record that satisfies four critical properties: consistency (schema-compliant representation), causality (temporal fidelity), fidelity (faithful to the agent’s internal and external behavior), and interoperability (analysis-ready, framework-agnostic).
This schema design builds on recent efforts in AI observability frameworks \cite{goyal2024llmguard} and structured introspection mechanisms for LLM agents \cite{rombaut2024watsoncognitiveobservabilityframework}, extending them with a formal schema for high-fidelity, surface-level trace capture. However, our schema uniquely emphasizes semantically enriched introspection in LLM agents, encompassing not just control flow and system I/O, but also the agent’s cognitive deliberations and interactions with external APIs and data stores.

\subsection{Surface Taxonomy and Extraction Procedure}

We operationalize the schema across three disjoint but composable surfaces of agent execution, each instrumented through techniques designed for transparent, non-intrusive logging.

\subsection{Operational Surface: Method-Level Execution Tracing}
The operational surface captures all explicit agent method calls, argument structures, return values, and execution timing. Through Python introspection and function wrapping, we automatically intercept all public methods on the agent class. Each method invocation produces a pair of events - start and complete - enriched with span-level metadata such as argument count, result type, and execution duration. Events are written to both structured JSONL logs and OpenTelemetry spans, preserving trace and span relationships for end-to-end visibility.

This approach is functionally equivalent to distributed tracing as used in service-oriented architectures \cite{sigelman2010dapper}, but adapted for fine-grained agent-level observability. All logs conform to a fixed schema to enable downstream consumption, and trace linkage ensures coherent propagation across multiple layers of abstraction.

\subsection{Cognitive Surface: LLM Interaction Introspection}
The cognitive surface is designed to capture the internal deliberations of the agent’s reasoning engine, primarily interactions with LLMs. These include raw prompts, completions, extracted reasoning chains (e.g., Chain-of-Thought), and confidence estimates. When supported by the LLM API (e.g., OpenAI or Anthropic), this surface also parses semi-structured outputs to extract \texttt{<thinking>} segments, step-by-step reasoning, and structured JSON fields such as plan or reflection.

Span metadata is derived from instrumented LLM API calls, and cognitive spans are nested within operational traces to maintain full trace context. 

Extraction relies on a set of generalizable strategies: marker-based pattern detection, XML tag parsing, and JSON field extraction. This design supports multiple reasoning formats and enables comparative analysis across different model outputs and prompt templates.

\begin{figure*}[t]
  \centering
  \includegraphics[width=0.95\linewidth]{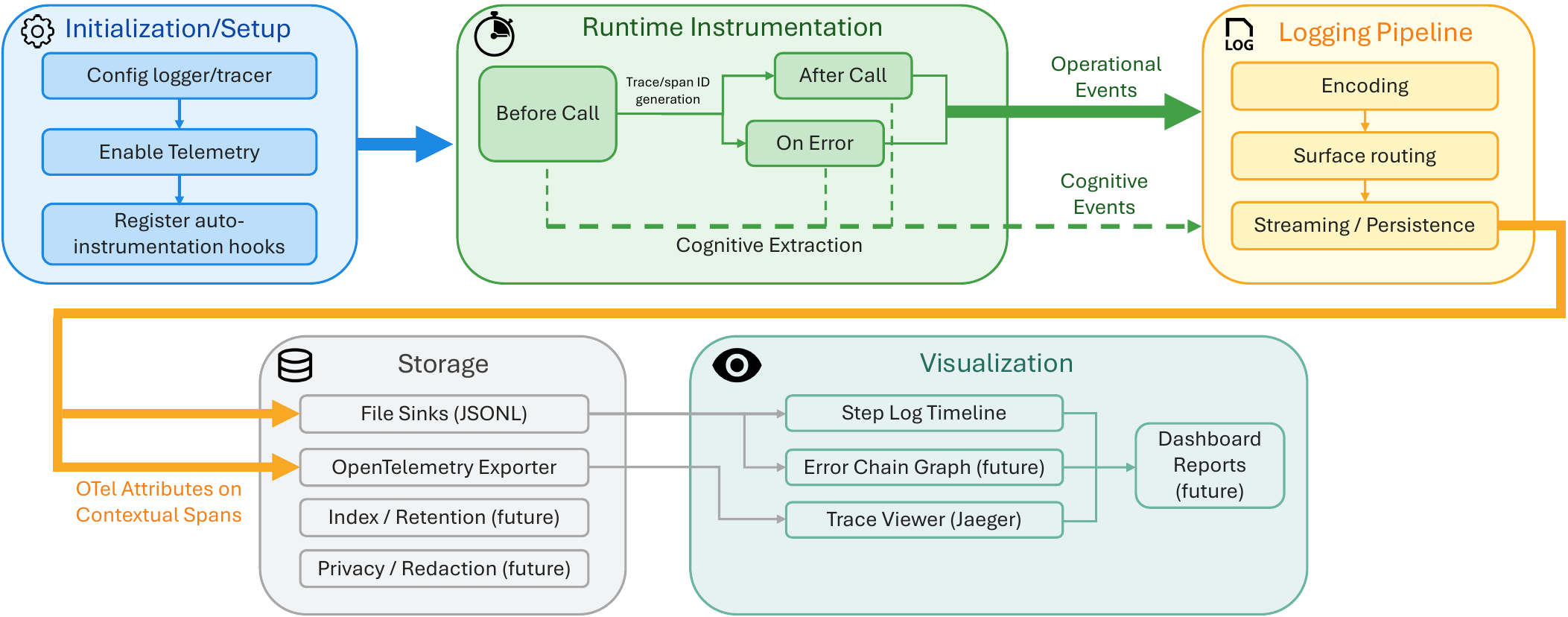}
  \caption{\textbf{AgentTrace system flow.} End-to-end runtime architecture showing initialization (logger setup, OpenTelemetry enablement, and auto-instrumentation hooks), runtime instrumentation with trace/span ID generation and cognitive extraction, logging pipeline for event encoding and routing, and downstream storage and visualization. Contextual spans are enriched with OpenTelemetry attributes via auto-instrumentation.}
  \label{fig:flowchart}
\end{figure*}

\subsection{Contextual Surface: External System I/O}
The contextual surface tracks all outbound agent interactions with external systems, including HTTP APIs, SQL/NoSQL databases, cache layers, vector stores, and file systems. Rather than requiring agent authors to manually log these operations, we leverage OpenTelemetry’s auto-instrumentation capabilities to monkey-patch standard libraries (e.g., requests, sqlalchemy, redis) at runtime. However, we also include the option for manual instrumentation in order to capture more granular and cohesive log structures.

Each contextual interaction produces a span enriched with resource-specific metadata: URLs and headers for HTTP, query structure and row counts for SQL, and key/value operations for cache or vector DBs. These events are stored exclusively as OTel spans to avoid redundancy with file-based logging. Temporal nesting under the same trace context enables causal analysis across layers of computation and I/O.

This surface bridges the agent’s cognitive operations and their environmental grounding, providing a unified view of how internal plans translate to external effects (cf. Paxton et al. 2023).

\subsection{Unified Representation and Trace Semantics}

All three surfaces emit logs that conform to a shared envelope schema. Each log event includes:

\begin{itemize}
  \item a UUID identifier
  \item surface type (cognitive, operational, or contextual)
  \item trace ID and span ID
  \item precise UTC timestamp
  \item event body (structured per surface)
\end{itemize}

\noindent Logs are stored in two complementary formats:

\begin{itemize}
    \item JSONL files (line-delimited JSON for offline inspection, streaming, or replay)
    \item OpenTelemetry spans (for real-time distributed tracing and integration with tools like Jaeger or Tempo)
\end{itemize}

This dual-path storage ensures both low-latency local debugging and scalable remote observability. Logs are append-only and schema-validated at write time to preserve consistency and support batch analytics and visualization.

\section{Implementation}
\label{sec:implementation}

To operationalize the proposed schema, we implemented AgentTrace as a modular runtime system that instantiates the three observability surfaces introduced in the Methodology section. This implementation translates the theoretical framework into concrete runtime mechanisms for capturing, structuring, and exporting agent logs.

\paragraph{Overview.}
AgentTrace is a lightweight Python package that (i) injects runtime instrumentation without modifying agent code, (ii) emits schema-consistent records across \emph{operational}, \emph{cognitive}, and \emph{contextual} surfaces, and (iii) exports telemetry to an OpenTelemetry (OTel) backend for distributed tracing. Our design goals are non-intrusiveness, low overhead, and graceful degradation (i.e., falling back to local logging when remote export fails) when external telemetry is unavailable.

\subsection{Python Module Layout and Initialization}
\label{sec:impl:module}

\textbf{AgentTrace} exposes a minimal API:
\begin{itemize}
  \item \texttt{init(...)} configures local sinks and toggles OTel export or auto-instrumentation.
  \item \texttt{instrument\_agent(obj, name, methods=None)} wraps selected public callables for runtime tracing.
  \item \texttt{ALogger} records surface-specific events and optionally exports to OTel.
\end{itemize}
Initialization loads configuration, prepares append-only JSONL outputs, and, when enabled, activates OTel auto-instrumentation for common I/O libraries.

\subsection{Instrumentation via Decorator Injection}
\label{sec:impl:instrumentation}

AgentTrace uses a runtime observer pattern. For each target method, it installs an in-place wrapper that:
\begin{enumerate}
  \item emits a \emph{start} event (method name, argument summary, timestamp),
  \item records a \emph{complete} event on success (duration, result summary), and
  \item records an \emph{error} event on exception while re-raising to preserve semantics.
\end{enumerate}
Wrappers preserve function signatures, and each event carries a fresh \texttt{span\_id} under a shared \texttt{trace\_id}. Span nesting and context propagation enforce temporal causality across reasoning, tool, and workflow events.

\paragraph{Cognitive trace extraction.}
When completions include a delimited reasoning segment, AgentTrace returns the cleaned answer and logs the segment as a \emph{cognitive} event; otherwise, results pass through unmodified.

\begin{algorithm}[tb]
\caption{AgentTrace Runtime Instrumentation Wrapper}
\label{alg:agenttrace-wrapper}
\textbf{Input}: agent instance $A$, name $n$, optional allowlist $\mathcal{M}$\\
\textbf{Output}: instrumented agent $A$

\begin{algorithmic}[1]
\FOR{each public method $m \in$ \texttt{select}$(A,\mathcal{M})$}
  \STATE let $f \leftarrow$ original implementation of $m$
  \STATE define wrapper $w$ with preserved signature
  \STATE \textbf{function} $w(\mathbf{x})$
    \STATE \quad $(\texttt{trace\_id}, \texttt{span\_id}) \leftarrow$ new IDs (or propagate)
    \STATE \quad \textsc{RecordOperational}(
      \texttt{status}=\texttt{start}, $n,m$,
      \texttt{args}=\texttt{summary}$(\mathbf{x})$
    )
    \STATE \quad $t_0 \leftarrow$ now()
    \STATE \quad \textbf{try}
      \STATE \qquad $y \leftarrow f(\mathbf{x})$
      \STATE \qquad $(y, \theta) \leftarrow$ \texttt{maybe\_extract\_cognitive}$(y)$
      \STATE \qquad \textbf{if} $\theta \neq \varnothing$ \textbf{then}
        \textsc{RecordCognitive}$(\theta)$
      \STATE \qquad \textsc{RecordOperational}(
        \texttt{dur}=now()$-t_0$,
        \texttt{status}=\texttt{complete},
        \texttt{result}=\texttt{summary}$(y)$
      )
      \STATE \qquad \textbf{return} $y$
    \STATE \quad \textbf{catch} exception $e$
      \STATE \qquad \textsc{RecordOperational}(
        \texttt{status}=\texttt{error},
        \texttt{dur}=now()$-t_0$,
        \texttt{err}=\texttt{repr}$(e)$
      )
      \STATE \qquad \textbf{rethrow} $e$
    \STATE \textbf{end function}
  \STATE replace $m$ on $A$ with $w$
\ENDFOR
\STATE \textbf{return} $A$
\end{algorithmic}
\end{algorithm}

\subsection{Log Schema and Local Sinks}
\label{sec:impl:schema}

All surfaces share a common envelope with identifiers, timestamps, agent name, \texttt{surface} $\in \{\text{operational},\text{cognitive},\text{contextual}\}$, level, \texttt{trace\_id}, and \texttt{span\_id}.
Surface payloads are concise: \textbf{Operational} includes method, status, duration, and result summary (optionally token or latency metadata); \textbf{Cognitive} stores thought, plan, and reflection excerpts with model and token counts; \textbf{Contextual} captures operation type, source, query or response summaries, and provenance. 
Crucially, contextual traces manage tool invocations and data access operations (reads and writes), linking agent reasoning with its external interactions. 
Operational and cognitive events are persisted to JSONL, while contextual events are primarily exported via OTel with optional file mirroring for offline workflows. 
All records are validated at emission time against the schema to maintain consistency.

\subsection{OpenTelemetry Export and Auto-Instrumentation}
\label{sec:impl:otel}

When enabled, AgentTrace converts each event into an OTel span and exports via a batch processor. Attributes are populated defensively, i.e., with type checks and safe conversions: scalars are set directly, structured values are JSON-stringified when necessary, and unknown objects fall back to string representations. Exporter failures gracefully degrade to local JSONL.

\paragraph{Contextual I/O capture.}
OTel auto-instrumentation patches common HTTP, database, and cache libraries to emit contextual spans (URLs, queries, status, counts, latencies) without manual logging. In tracing UIs (e.g., Jaeger), method-level operational spans appear alongside contextual spans, providing end-to-end visibility complementary to local files.

\subsection{Engineering Considerations}
\label{sec:impl:eng}

\textbf{Non-intrusiveness.} Decorators enable tracing without changing agent logic.\\
\textbf{Low overhead.} Typical success paths emit two events per call; export is batched asynchronously.\\
\textbf{Robustness.} Serialization and export are defensive; failures never block execution.\\
\textbf{Composability.} A stable, analysis-ready schema supports JSONL and OTel; contextual I/O is auto-instrumented with optional file mirroring.

\section{Conclusion}

In this paper, we present AgentTrace, a research framework that establishes the first open standard for structured agent logging through a schema-based protocol spanning cognitive, operational, and contextual traces. By transforming logging into a semantically rich, introspectable substrate, AgentTrace elevates observability from an engineering utility to a core enabler of agent safety, reproducibility, and accountability. This design closes critical gaps in existing observability systems by enabling fine-grained debugging, reliable failure attribution, and transparent governance of LLM-based agents.

Looking ahead, the structured and interpretable traces generated by AgentTrace pave the way for security and evaluation research. They allow for dynamic threat modeling, real-time risk detection, and post-hoc forensic analysis of adversarial or misaligned behaviors. Beyond security, these logs provide the groundwork for agent evaluation, enabling new metrics for reasoning stability, goal fidelity, and cross-agent behavioral benchmarking.


\begin{thebibliography}{10}

\bibitem{dong2024agentopsenablingobservabilityllm}
Dong, L.; Lu, Q.; and Zhu, L.
\newblock 2024.
\newblock {AgentOps: Enabling Observability of LLM Agents}.
\newblock {\em arXiv preprint arXiv:2411.05285}.

\bibitem{goyal2024llmguard}
Goyal, S.; Hira, M.; Mishra, S.; Goyal, S.; Goel, A.; Dadu, N.; DB, K.; Mehta,
  S.; and Madaan, N.
\newblock 2024.
\newblock LLMGuard: Guarding Against Unsafe LLM Behavior.
\newblock In {\em Proceedings of the AAAI Conference on Artificial
  Intelligence (AAAI-24)},  23790--23792.
\newblock Demonstration Track.

\bibitem{kim2025llmspursueagenticinterpretability}
Kim, B.; Hewitt, J.; Nanda, N.; Fiedel, N.; and Tafjord, O.
\newblock 2025.
\newblock Because we have LLMs, we Can and Should Pursue Agentic
  Interpretability.

\bibitem{liu2024agentpatterns}
Liu, Y.; Lo, S.; Lu, Q.; Zhu, L.; Zhao, D.; Xu, X.; Harrer, S.; and Whittle,
  J.
\newblock 2024.
\newblock Agent Design Pattern Catalogue: A Collection of Architectural
  Patterns for Foundation Model Based Agents.
\newblock {\em arXiv preprint arXiv:2405.10467}.

\bibitem{rombaut2024watsoncognitiveobservabilityframework}
Rombaut, B.; Masoumzadeh, S.; Vasilevski, K.; Lin, D.; and Hassan, A.~E.
\newblock 2024.
\newblock Watson: A Cognitive Observability Framework for the Reasoning of
  LLM-Powered Agents.

\bibitem{rorseth2025ladybug}
Rorseth, J.; Godfrey, P.; Golab, L.; Srivastava, D.; and Szlichta, J.
\newblock 2025.
\newblock {LADYBUG: an {LLM} Agent {DeBUG}ger for data-driven applications}.
\newblock In {\em Proceedings of the 28th International Conference on
  Extending Database Technology (EDBT)},  1082--1085.
\newblock Demonstration Paper.

\bibitem{schick2023toolformer}
Schick, T.; Dwivedi-Yu, J.; Dess\`{i}, R.; Raileanu, R.; Lomeli, M.; Hambro,
  E.; Zettlemoyer, L.; Cancedda, N.; and Scialom, T.
\newblock 2023.
\newblock Toolformer: Language Models Can Teach Themselves to Use Tools.
\newblock In {\em Advances in Neural Information Processing Systems 36
  (NeurIPS 2023)}.

\bibitem{sigelman2010dapper}
Sigelman, B.~H.; Barroso, L.~A.; Burrows, M.; Stephenson, P.; Plakal, M.;
  Beaver, D.; Jaspan, S.; and Shanbhag, C.
\newblock 2010.
\newblock Dapper, a Large-Scale Distributed Systems Tracing Infrastructure.
\newblock Google Technical Report.

\bibitem{weng2023autonomous}
Weng, L.
\newblock 2023.
\newblock LLM Powered Autonomous Agents.
\newblock Online blog post,
  {\url{https://lilianweng.github.io/posts/2023-06-23-agent/}}.

\bibitem{Zheng_2025}
Zheng, Y.; Hu, Y.; Yu, T.; and Quinn, A.
\newblock 2025.
\newblock AgentSight: System-Level Observability for AI Agents Using eBPF.
\newblock In {\em Proceedings of the 4th Workshop on Practical Adoption
  Challenges of ML for Systems}, SOSP '25,  110--115.
\newblock ACM.

\bibitem{zou2023universal}
Zou, A.; Wang, Z.; Carlini, N.; Nasr, M.; Kolter, J.~Z.; and Fredrikson, M.
\newblock 2023.
\newblock Universal and Transferable Adversarial Attacks on Aligned Language
  Models.
\newblock {\em CoRR}  abs/2307.15043.

\end{thebibliography}
\end{document}